# Cervical Optical Coherence Tomography Image Classification Based on Contrastive Self-Supervised Texture Learning

Kaiyi Chen, Qingbin Wang, and Yutao Ma*

School of Computer Science, Wuhan University, Wuhan 430072, China

{chris, qbwang, ytma}@whu.edu.cn

*Corresponding author (ORCID): 0000-0003-4239-2009

*Abstract—Background:* Cervical cancer seriously affects the health of the female reproductive system. Optical coherence tomography (OCT) emerged as a non-invasive, high-resolution imaging technology for cervical disease detection. However, OCT image annotation is knowledge-intensive and time-consuming, which impedes the training process of deep-learning-based classification models. *Purpose:* This study aims to develop a computer-aided diagnosis (CADx) approach to classifying *in-vivo* cervical OCT images based on self-supervised learning. *Methods:* In addition to high-level semantic features extracted by a convolutional neural network (CNN), the proposed CADx approach designs a contrastive texture learning (CTL) strategy to leverage unlabeled cervical OCT images' texture features. We conducted ten-fold cross-validation on the OCT image dataset from a multi-center clinical study on 733 patients from China. *Results:* In a binary classification task for detecting high-risk diseases, including high-grade squamous intraepithelial lesion and cervical cancer, our method achieved an area-under-the-curve value of $0.9798 \pm 0.0157$ with a sensitivity of $91.17 \pm 4.99\%$ and a specificity of $93.96 \pm 4.72\%$ for OCT image patches; also, it outperformed two out of four medical experts on the test set. Furthermore, our method achieved a 91.53% sensitivity and 97.37% specificity on an external validation dataset containing 287 3D OCT volumes from 118 Chinese patients in a new hospital using a cross-shaped threshold voting strategy. *Conclusions:* The proposed contrastive-learning-based CADx method outperformed the end-to-end CNN models and provided better interpretability based on texture features, which holds great potential to be used in the clinical protocol of "see-and-treat."

## 1 Introduction

Cervical cancer ranked fourth for both incidence and mortality among females in 2018.[1] About 570,000 new cases and 311,000 deaths from cervical cancer occurred in 2018,[2] and over 80% of the new cases and deaths occurred in poor and developing countries. It has been recognized that almost all cases of cervical cancer are caused mainly by the human papillomavirus (HPV). Fortunately, the World Health Organization (WHO) has treated cervical cancer as a public health problem and adopted specific strategies to accelerate the elimination of cervical cancer, including effective HPV vaccination, cervical screening, and timely treatment of precancerous lesions.[3]

There are several standard clinical screening methods for cervical disease detection, such as the HPV test, thin-prep cytologic test (TCT), and colposcopy. However, these methods have their respective disadvantages.[4,5] For example, the HPV test has a relatively high false-positive rate, and both TCT and colposcopy have a reasonably high chance of missed diagnoses of high-risk cervical lesions. The cervical biopsy with histopathological confirmation is the gold standard to diagnose cervical lesions. However, it is invasive and time-consuming; besides, the rate of missed diagnosis is relatively high due to biopsy sampling errors. Therefore, developing rapid, non-invasive, effective, and intelligent screening and diagnostic approaches[6] is necessary to achieve the WHO 2030 targets.

Optical coherence tomography (OCT)[7] is an emerging non-invasive three-dimensional (3D) imaging technology developed in the early 1990s. An OCT system uses light waves to obtain high-resolution, cross-sectional images in biological systems. Currently, OCT has widely been applied in ophthalmology, cardiology, and other fields of clinical medicine. Moreover, recent studies of OCT imaging on the cervix[5, 8-10] have shown that this technology can capture histomorphological features of cervical tissue *ex vivo* and *in vivo* and thus enables pathologists or gynecologists to make a precise diagnosis. Unfortunately, most pathologists and gynecologists do not get familiar with cervical OCT images, implying a sharp learning curve. Therefore, we need to develop intelligent computer-aided diagnosis (CADx) approaches to help them analyze cervical OCT images more efficiently.

In recent years, deep-learning-based CADx approaches for medical images with different modalities have achieved better results than traditional machine learning algorithms. The same is true in analyzing cervical screening images, such as cytology and colposcopy images. For example, Almubarak *et al.*[11] exploited multi-scale features to classify histology images from cervical tissue samples on a small-scale dataset and achieved an accuracy of 77.25%. Bhargava *et al.*[12] employed support vector machine (SVM), k-nearest neighbor, and artificial neural network (ANN) models to train Pap smear image classifiers using hand-craft features. They found that the ANN-based classifier performed best in the test on a small-scale dataset. Kudva *et al.*[13] presented a hybrid transfer learning approach for cervical cancer screening using two convolutional neural networks (CNNs): AlexNet and VGG-16. Priya *et al.*[14] proposed a heuristic and ANN-based classification model for early cervical cancer screening using linear SVM. Alyafeai *et al.*[15] proposed a fully automated pipeline for cervical cancer classification of cervigram images, consisting of two pre-trained deep learning models. Li *et al.*[16] proposed a deep learning framework for identifying cervical intraepithelial neoplasia (CIN) and cervical cancer using time-lapsed colposcopic images. They tested the proposed framework's performance and found it comparable to an in-service colposcopist. **Table I** summarizes the classification results of selected recent work for analyzing cytology and colposcopy images. However, none of the above utilizes the cutting-edge *in-vivo* OCT imaging technique due to the lack of licensed OCT devices and publicly available cervical OCT datasets.

Training deep-learning-based models for medical image analysis requires large amounts of labeled image data. Due to insufficient high-quality training data, many CADx approaches based on deep learning cannot reach the desired performance level, especially when dealing with new image datasets.[23] Medical image annotation, a specialized, labor-intensive, and time-consuming job, has become a bottleneck of model training. OCT image annotation is also an expensive procedure in labor and time costs. Furthermore, few medical experts are familiar with this new imaging technology. Therefore, the challenge we aim to tackle is to acquire sufficient labeled cervical OCT images. Self-supervised learning (SSL) has been proposed to leverage unlabeled data without human-annotated supervision. In recent years, the SSL methods have achieved great success on images and texts as an unsupervised learning approach for representation learning without manual data labels. In this study, SSL is thus a proper solution to train deep-learning-based OCT image classification models for cervical cancer screening and diagnosis.

TABLE I. SUMMARY OF SELECTED CLASSIFICATION MODELS FOR CERVICAL SCREENING IMAGES

| Image type | Literature | Model | Performance | Data size |
|---|---|---|---|---|
| Pap smear test | Ref.[12] | deep learning | accuracy: 95.5% | 66 images |
| Pap smear test | Ref.[21] | deep learning + SVM | accuracy: 98.32% and 97.87% for two datasets | dataset1: 917 images; dataset2: 4,049 images |
| colposcopy | Ref.[13] | deep learning | accuracy: 91.46% | 1,644 images |
| colposcopy | Ref.[16] | deep learning | accuracy: 78.33% | 7,668 patients |
| colposcopy | Ref.[17] | SVM | sensitivity: 81.3% and specificity: 78.6% for detecting CIN+ | 134 patients |
| colposcopy | Ref.[18] | deep learning | accuracy: 81.35% | 800 images |
| colposcopy | Ref.[19] | deep learning | accuracy: 73.08%; AUC: 0.75 for detecting CIN2/CIN3+ | 1,709 patients |
| colposcopy | Ref.[20] | deep learning | accuracy: 85.5%; AUC: 0.909 for detecting HSIL+ | 1,400 patients |
| cytology | Ref.[21] | deep learning + SVM | accuracy: 99.47% | 963 images |
| cytology | Ref.[22] | deep learning | accuracy: 88.84%; F-score: 59.96% | 14,432 image patches |

Recently, SSL was also used to alleviate the vital requirement of data annotation for medical images. For example, Zhu *et al.*[24] proposed a pretext task that sorts slices extracted from 3D computed tomography (CT) and magnetic resonance imaging (MRI) volumes and fine-tuned a 3D neural network model on a small amount of annotated data. Taleb *et al.*[25] conducted a similar study and released their implementations as open-source libraries. These 3D SSL methods can utilize the full 3D spatial context and learn translation and rotation invariant features from the original 3D volumetric scans, improving data efficiency and performance on different downstream tasks. Besides, Chen *et al.*[26] proposed a context restoration strategy to learn useful semantic features of three different types of medical images. Li *et al.*[27] presented an SSL method that learns rotation-related and rotation-invariant features for retinal disease diagnosis. Their abilities to explore the invariant property of unlabeled medical images for classification can improve model generalization. However, most existing SSL methods neglect the local texture information recognized to be beneficial for diagnosing gray-scale images, such as CT, MRI, and OCT.

To tackle the challenge of expensive OCT image annotation, we present a self-supervised contrastive texture learning strategy in this study. Specifically, it can leverage the local texture information and high-level semantic features extracted by a CNN model to automatically learn visual representations of OCT images for different cervical diseases. After the pre-training process ends, we need only a small-scale dataset of labeled cervical OCT images to fine-tune the CNN model for classification, reducing the burden on image annotation. Also, we develop a CADx approach to assist gynecologists in efficiently analyzing cervical OCT images in clinical practice. The CADx approach consists of an image classification model pre-trained by the proposed SSL strategy and a feature visualization module highlighting possible lesions for gynecologists. In brief, the main contributions of this study are three-fold.

- We first introduce SSL to *in-vivo* cervical OCT image classification and specify a contrastive texture learning (CTL) strategy to learn semantic and texture features from unlabeled cervical OCT images effectively, providing better explainable evidence for gynecologists to make a proper diagnosis.
- We develop a deep-learning-based CADx approach for cervical cancer screening and diagnosis using OCT imaging. In addition to accurate image classification, the CADx approach can highlight possible lesions automatically according to the learned histomorphological and local texture features.
- We experiment on a cervical OCT image dataset from a multi-center clinical study. The proposed CTL strategy outperforms state-of-the-art SSL methods regarding evaluation metrics, and the CNN-based image classification model pre-trained by the CTL strategy can match four medical experts in detecting high-risk cervical lesions.

The rest of this article is organized as follows. Section 2 introduces the CADx method and OCT image dataset used in this study, and Section 3 presents the experimental results. Potential threats to the validity of this study are discussed in Section 4. Finally, Section 5 concludes this article.

## 2 MATERIALS AND METHODS

### 2.1 Data Collection

The OCT image dataset was collected from a multi-center clinical study using OCT to evaluate cervical lesions *in vivo*.[5] Seven hundred thirty-three gynecological outpatients were recruited in five hospitals in China from August 2017 to December 2019. The standard for patient recruitment is that one or both cervical screening results (i.e., HPV and TCT results) were positive. Each of the patients recruited for this multi-center clinical study was inspected with OCT and received colposcopy-directed cervical biopsy. **Table II** presents the demographic information of all the patients.[5]

TABLE II.  DEMOGRAPHIC INFORMATION OF PATIENTS

| Hospital | #Patients | Age (mean±std) | HPV results | TCT results |
| --- | --- | --- | --- | --- |
| The Third Affiliated Hospital of Zhengzhou University | 350 | 38.67±9.86 | Positive: 281 Negative: 31 Untested: 38 | Positive: 215 Negative: 63 Untested: 72 |
| Liaoning Cancer Hospital and Institute | 227 | 44.08±8.38 | Positive: 138 Negative: 84 Untested: 5 | Positive: 69 Negative: 134 Untested: 24 |
| Puyang Oilfield General Hospital | 59 | 43.04±8.06 | Positive: 39 Negative: 3 Untested: 17 | Positive: 39 Negative: 12 Untested: 8 |
| Luohe Central Hospital | 57 | 40.37±10.05 | Positive: 49 Negative: 8 Untested: 0 | Positive: 36 Negative: 21 Untested: 0 |
| Zhengzhou Jinshui District General Hospital | 40 | 39.03±12.47 | Positive: 36 Negative: 2 Untested: 2 | Positive: 9 Negative: 27 Untested: 4 |
| Overall | 733 | 40.85±9.79 | Positive: 543 Negative: 128 Untested: 62 | Positive: 368 Negative: 257 Untested: 108 |

A 3D OCT volume from the point of a 12-hour clock on the cervix contains 10 or 20 frames, each corresponding to an *in-vivo* cervical OCT image. The class label of each 3D OCT volume was set according to the histopathology-confirmed diagnosis for the corresponding biopsy specimen. Due to biopsy sampling errors and image quality issues, we selected 1,256 3D OCT volumes from 699 patients, matching the corresponding hematoxylin and eosin (H&E) stained images. Because normal cervical tissues and local lesions often appear together in an OCT image (i.e., one frame of an OCT volume), pathologists then extracted patches of 600×600 pixels that contain the lesion area from each selected OCT image according to the corresponding H&E stained image. As with previous studies,[5, 6, 8] this study defines mild inflammation (MI), ectropion (EP), and cyst (CY) as low-risk or negative, and high-grade squamous intraepithelial lesion (HSIL) (including CIN II and CIN III) and cervical cancer (CC) as high-risk or positive. **Table III** shows the statistics of the cervical OCT image dataset.

TABLE III. STATISTICS OF EXPERIMENTAL DATASET

| Size | MI | EP | CY | HSIL | CC | Total |
|---|---|---|---|---|---|---|
| #Patients | 239 | 126 | 99 | 161 | 74 | 699 |
| #Volumes | 363 | 195 | 153 | 166 | 379 | 1,256 |
| #Patches | 3,172 | 2,464 | 2,067 | 5,539 | 731 | 13,973 |

Besides, we collected 287 3D OCT volumes from 118 outpatients (2.43 points per patient) in the Second Xiangya Hospital of Central South University as an external validation dataset to evaluate our CADx approach's generality, approved by the Ethics Committees of the hospital. Each OCT volume has a specific label from histopathology or radiology in the same hospital. The 118 patients (or their legal guardians) signed informed consent.

2.2 Model Construction and Training

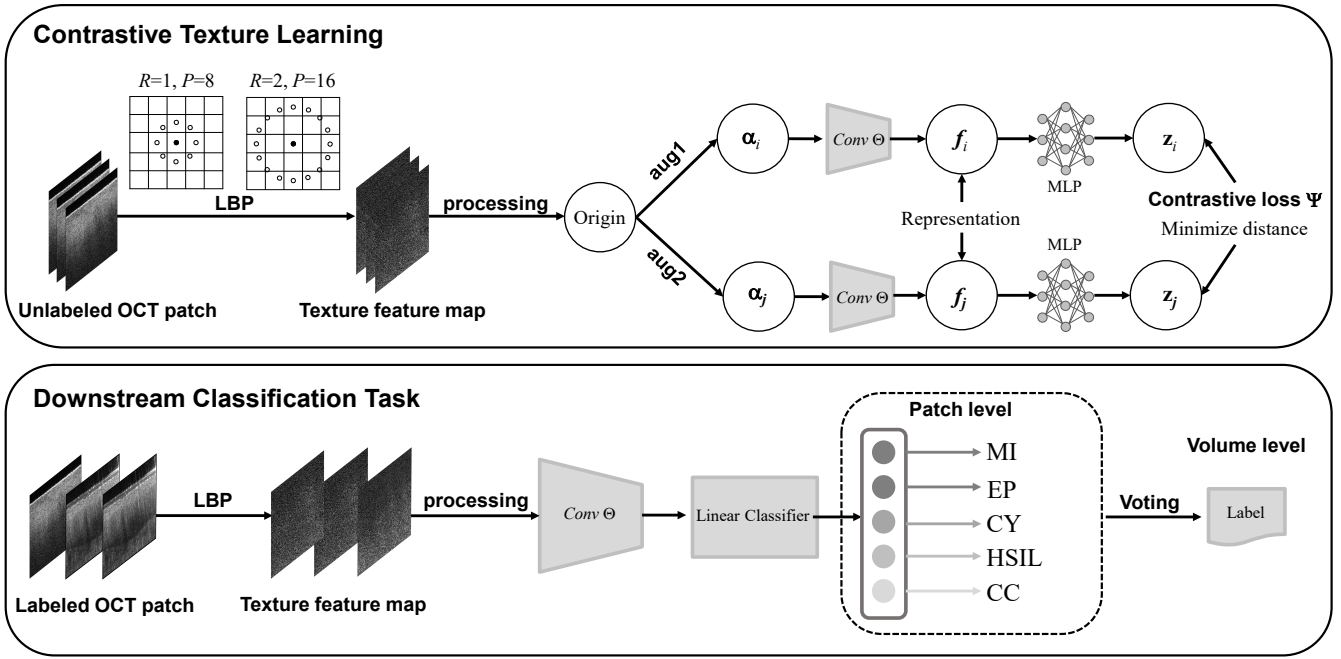

Fig. 1. The construction and training process of the classification model based on SSL.

The core of our CADx approach is a multi-granularity OCT image classification model for patches, images, and 3D volumes. This classification model is structured and trained in a self-supervised manner. SSL usually has two steps. First, it defines a specific pretext task to pre-train a model on a large-scale dataset of unlabeled data. The pretext task aims to learn representations for downstream tasks. Second, the pre-trained model is fine-tuned on a small-scale annotated dataset for specific downstream tasks. **Figure 1** presents an overview of the classification model's construction and training process.

*Pretext task definition*. We specify the pretext task for this study by CTL, which aims to build visual representations by learning to encode the similarity or difference between two texture feature maps. We first extract texture features from an original input image because the texture features of gray-scale medical images are particularly prominent at the pixel level. Here, we do not apply image augmentation to source OCT images for high efficiency in the pipeline. For more details of the texture extraction algorithm, please refer to Subsection 2.3. Then, we deal with texture feature maps by the standard min-max normalization method to avoid the local extremum. Finally, we generate a pair of new samples for each normalized texture feature map (denoted by the notation "Origin" in **Figure 1**) using data augmentation, such as random rotation and the vertical and horizontal flip augmentation, to provide richer texture features for CTL.

Due to the advantages of CNNs in image analysis, we also model the CTL strategy by the CNN architecture. Suppose we build a CNN network denoted as *Conv* $\Theta$. Then, normalized texture feature maps with data augmentation are fed into the network to generate feature maps. For more details of the CTL loss function, please refer to Subsection 2.4. In the pretext task, we employ a multilayer perceptron (MLP) to replace the original fully connected layer, called the projection head.[28] Thus, when performing the downstream classification task, the feature maps obtained by the network will be more generalized rather than biased towards distinguishing feature distributions of different classes of input images. Moreover, we use the batch normalization method for each CNN layer to avoid the internal covariate shift.

*Downstream task execution*. We fine-tune the pre-trained network *Conv* $\Theta$ for the downstream classification task using a small amount of labeled data. Specifically, we replace the last fully connected layer of the network (i.e., the MLP) with a global average pooling (GAP) layer to meet the size requirement and reduce the computation load. Besides, our CADx method can highlight texture features of cervical lesions learned from cervical OCT images to interpret diagnoses better. For more details of image classification, please refer to Subsection 2.5.

2.3 Texture Extraction

Ojala *et al.*[29] proposed a straightforward yet efficient approach to extract rotation invariant texture features from gray-scale images with local binary patterns (LBPs). Given an input image $G$, we characterize the spatial structure of local texture features by defining $LBP_{P,R,G}$ as

$$LBP_{P,R,G} = \sum_{p=0}^{P-1} I_{[(g_p-g_c \geq 0)]}(g_p - g_c)2^p, \qquad (1)$$

where $I(\cdot)$ is an indicator function, $g_c$ is the gray value of the center pixel of the local neighborhood, $g_p$ is the gray value of an equally spaced pixel on a circle of radius $R$ ($R > 0$), and $P$ is the number of pixels on the circle.

The LBP part in **Figure 1** illustrates circularly symmetric neighbor sets for different combinations of $P$ and $R$. If the center pixel corresponding to $g_c$ is located at the origin $(0,0)$, the coordinates of a pixel corresponding to $g_p$ are given by $(-R\sin(2\pi p/P), R\cos(2\pi p/P))$. To maintain rotation invariance, we then define a unique identifier for a rotation-invariant LBP as

$$LBP_{P,R,G}^{ri} = \min\{ROR(LBP_{R,R,G}, i) \mid i = 0,1, \dots, P-1\}, \quad (2)$$

where $ROR(\xi, i)$ performs shifting the binary bits of integer $\xi$ rightward for $i$ times. Empirical studies suggest that small $R$ and $P$ values often lower image classification performance. According to an ablation experiment (see Subsection 4.1), we set $R = 4$ and $P = 32$ in this study to obtain different cervical OCT image texture features.

2.4 CTL Loss Function

Contrastive learning is one of the most commonly-used pretext tasks in SSL. Its goal is to learn the similarity and the difference between various classes. Therefore, contrastive learning has widely been applied to build pre-training models using massive unlabeled data. Given a randomly sampled mini-batch of $B$ input samples (i.e., texture feature maps extracted by LBP), we generate $2 \times B$ new samples using data augmentation. For each positive pair $(i, j)$ of augmented samples from the same original input, we treat the remaining $2 \times (B - 1)$ augmented samples as negative samples. As shown in the contrastive loss part in **Figure 1**, we define the CTL loss function $\zeta$ of pair $(i, j)$ as

$$\zeta_{i,j} = -\log \frac{\exp(cosd((\Theta(\alpha_i), \Theta(\alpha_j)))/\tau)}{\sum_{k=1}^{2B} I_{[k \neq i]} \exp(cosd((\Theta(\alpha_i), \Theta(\alpha_k)))/\tau)}, \quad (3)$$

where $\Theta(\cdot)$ represents the network's output, $cosd(\mathbf{a}, \mathbf{b}) = \mathbf{a}^T \cdot \mathbf{b}/\|\mathbf{a}\|\|\mathbf{b}\|$ represents the cosine distance between vectors $\mathbf{a}$ and $\mathbf{b}$, and $\tau$ is a temperature parameter. Then, the whole loss value $\Psi$ of a mini-batch is defined as

$$\Psi = \frac{1}{2B} \sum_{k=1}^{B} (\zeta_{2k-1,2k} + \zeta_{2k,2k-1}). \quad (4)$$

2.5 Image Classification

The MLP used for the pretext task is replaced by a linear classifier to obtain more generalized feature information in the downstream task. Specifically, this linear classifier comprises a fully connected layer followed by a softmax function. The network is fine-tuned on a small amount of annotated cervical OCT image patches by leveraging the features learned in the pre-training process. The cross-entropy loss function for classification is formulated in **Equation (5)**. After the fine-tuning process ends, it outputs the corresponding probability distributions for the five categories (i.e., MI, EP, CY, HSIL, and CC) and generates a specific label for an input patch. The likelihood that the input patch belongs to a binary class (i.e., high-risk and low-risk) is inferred by summing up the probabilities over all the categories within the general class.

$$\Phi = \frac{1}{N}\sum_i \Phi_i = -\frac{1}{N}\sum_i \sum_{c=1}^{M} y_{ic} \log p_{ic}, \qquad (5)$$

where $N$ is the number of samples, $M$ is the number of categories, $y_{ic} = I(i = c)$ that returns zero or one, and $p_{ic}$ represents the probability that sample $i$ belongs to category $c$.

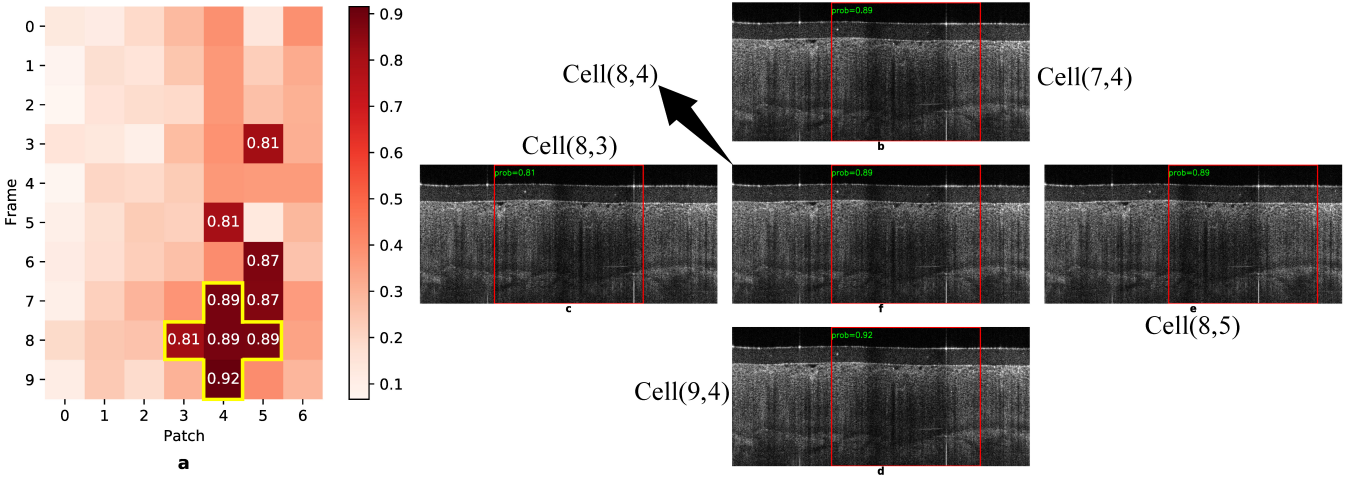

Fig. 2. An illustration of the cross-shaped threshold voting strategy.

For an OCT volume consisting of $m$ frames, we use slicing windows to extract $n$ patches from the same frame. As a result, we can obtain a $m \times n$ matrix of predictions for patches. Furthermore, according to the clinical experience of medical experts, in this study, we design an explainable voting strategy, called cross-shaped threshold voting, to aggregate patch-level predictions into a single result for the whole OCT volume. The basic idea of this strategy is to detect consecutive high-risk patches in the cross-sectional and axial directions, which may look like a cross shape. Moreover, we define a threshold value of 0.8 for high-risk (or positive) to better balance the false-positive and false-negative rates. **Figure 2** illustrates the cross-shaped threshold voting strategy. Our method detected five high-risk patches in this volume, including the fifth patch in the eighth, ninth, and tenth frames and the fourth, fifth, and sixth patches in the ninth frame.

2.6 Feature Visualization

Because the classification model is built using the CNN architecture, we also provide volume- and patch-level feature visualization to assist gynecologists in detecting cervical lesions. For example, for an OCT volume, we present a heat matrix of size $m \times n$ to help localize a possible lesion (see the left part in **Figure 2**) from a 3D perspective. Besides, we use the class activation map (CAM)[30] that generates a heat map highlighting regions most relevant to the corresponding category based on OCT images' texture features to interpret patch-level diagnoses better.

2.7 Experiment Setups

To test the performance of our classification model, we designed two groups of experiments: a machine-machine comparison with different deep learning algorithms and a human-machine comparison with medical experts.

***Data partition***. The experimental dataset was randomly divided into a training set and a test set, with an 80:20 split based on the patient identification number, for the two comparison experiments. We applied the ten-fold cross-validation method to the training set for the machine-machine comparison experiment. In addition, we used an external validation dataset to evaluate the generality of the proposed CADx method. The training set was then divided into two subsets with an 80:20 split in the human-machine comparison experiment. Therefore, 20% of label-free OCT patches in the training set were used for SSL from scratch and the remaining 80% for the downstream classification task. To avoid the overfitting problem, OCT images from the same patient cannot appear in the training and test sets. Besides, we used the oversampling technique to address the class imbalance problem for high-risk (or positive) classes.

***Baselines for comparison***. We employed three standard CNN models (i.e., VGG19,[31] ResNet-50, and Resnet-101)[32] as the backbone networks in the machine-machine comparison experiment. In addition, we trained classification models with different initialization strategies to compare CTL with competitive baselines, including state-of-the-art SSL strategies, namely context restoration (CR),[26] SimCLR,[28] and SimSiam,[33] and the supervised learning method for training a model from scratch (denoted as *random initialization*). Moreover, we used EfficientNet-B7[34] pre-trained on the ImageNet dataset as a supervised pre-training baseline method. Besides, four medical experts from the Third Affiliated Hospital of Zhengzhou University, including one pathologist, one gynecologist, and two radiographers, were involved in the human-machine comparison experiment as anonymous investigators. All of them have learned and used OCT for more than one year. The pathologist is an associate chief physician skilled in the pathological diagnosis of gynecological and breast tumors. The gynecologist is a chief physician skilled in diagnosing and treating various malignant tumors in gynecology. The two radiographers are attending physicians skilled in the imaging diagnosis of obstetrics gynecological and breast diseases.

***Evaluation metrics***. The *accuracy* and *micro-F1 score* metrics were used in the five-class classification task. In addition, two frequently-used metrics, i.e., *sensitivity* and *specificity*, were used to evaluate the results in the binary classification task. Also, the area under the ROC curve (AUC) was employed to evaluate a given model's overall performance in the binary classification task. Finally, the machine-machine comparison results were made in terms of the average metric value of the ten-fold cross-validation results.

$$accuracy = \frac{TP+TN}{TP+FP+TN+FN}, \qquad (6)$$

$$\text{micro-}F1 = \frac{2 \cdot precision_{\text{micro}} \cdot recall_{\text{mirco}}}{precision_{\text{micro}} + recall_{\text{mirco}}}, \qquad (7)$$

$$precision_{\text{micro}} = \frac{\sum_{i=1}^{n} TP_i}{\sum_{i=1}^{n} TP_i + \sum_{i=1}^{n} FP_i}, \qquad (8)$$

$$recall_{\text{micro}} = \frac{\sum_{i=1}^{n} TP_i}{\sum_{i=1}^{n} TP_i + \sum_{i=1}^{n} FN_i}, \qquad (9)$$

$$sensitivity = TP/(TP + FN), \qquad (10)$$

$$specificity = TN/(TN + FP), \qquad (11)$$

where $TP$, $FP$, $TN$, and $FN$ represent true positives, false positives, true negatives, and false negatives, respectively.

*Environment and Configuration.* The experiments were conducted on an Asus X99-E WS/USB 3.1 workstation with an Intel Xeon E5-2620 v4 (2.10 GHz) and four graphics processing units (NVIDIA GeForce RTX 2080, 8 GB). The operating system in the workstation was Ubuntu 16.04, and the programming language was Python 3.6.2. Moreover, we used Pytorch 1.4.0 and sckit-learn 0.24.1 as the basic libraries of deep learning.

The backbone networks were used based on standard VGG19, ResNet-50, and Resnet-101 models. The MLP was set as a two-layer densely connected network, with the first layer containing 512 neurons and 128 neurons in the second layer. In the entire pre-training process of our classification model, we adopted adaptive gradient descent (Adam)[35] as the optimizer of the network $Conv\ \Theta$ with a 1e-2 learning rate and a 1e-6 weight decay. We adopted stochastic gradient descent (SGD)[36] as an optimizer with a 5e-3 learning rate and a 0.9 momentum for linear image classification. In all the experiments, the batch size was set to 32, and the number of epochs was set to 150. For more details of the implementation of our model, please refer to the source code publicly available at https://github.com/whuchris/MIA-CTL.

We used the standard zero-score method to normalize the input OCT images for the baseline models. We employed the same data augmentation techniques mentioned in the original article of CR.[26] In the process of context restoration, two patches in an image were randomly selected and swapped ten times. The $\ell_2$ loss function was optimized by Adam with $\beta_1 = 0.9$, $\beta_1 = 0.999$, and $\epsilon$ = 1e-8. Batch normalization was used in all CNNs. SimCLR used the default settings specified in its original paper[28] except that the learning rate was scalably adjusted according to the small batch size in the experiments. The SimSiam strategy used the same data augmentation method as SimCLR and CTL. We also utilized SGD with a base learning rate of 0.05 for pre-training to minimize the negative cosine similarity function. The weight decay was set to 1e-4, and the SGD momentum was set to 0.9. Note that for the downstream task of this study, we replaced the reconstruction part of CR, the projection head of SimCLR, and the projection head and prediction head of SimSiam, with the same MLP used by CTL.

## 3 RESULTS

### 3.1 Machine-machine comparison for OCT patches

In the five-class classification task, CTL on ResNet-101 achieved the best accuracy of 85.38±5.51% (mean±std), followed by CTL on VGG19 with an 84.77±3.96 accuracy. CTL on ResNet-101 also achieved the best performance in terms of F1-score. Compared with the state-of-the-art SimSiam, the accuracy (F1-score) values of CTL on ResNet-101 and VGG19 were increased by 1.41% (0.42%) and 2.22% (0.57%), respectively. According to a Wilcoxon signed-rank test[37] at the significance level of 0.05, there is a significant difference between CTL and SimSiam. This result indicates that introducing the texture features of cervical OCT images to contrastive learning can make classification results better. Moreover, the accuracy (F1-score) values of CTL on VGG19, ResNet-50, and ResNet-101 networks were 3.66% (4.05%), 5.02% (3.77%), and 2.99% (1.64%) higher than those of supervised learning CNN models without pre-training, respectively. Similarly, the models pre-trained by other contrastive learning strategies and the ImageNet dataset performed better than those trained from scratch in most cases. This result suggests that pre-training is indeed helpful to improve image classification performance in case of insufficient labeled data.

TABLE IV. COMPARISON AMONG CNN-BASED CLASSIFICATION MODELS (MEAN±STD)

| Model | Type | Init | Five-class Classification | | Binary Classification | | | |
|---|---|---|---|---|---|---|---|---|
| | | | *Accuracy* (%) | *F1-score* (%) | *Accuracy* (%) | *Sensitivity* (%) | *Specificity* (%) | AUC |
| VGG19 | Supervised | Random | 81.11±5.96 | 87.11±5.12 | 88.02±1.70 | 86.61±6.30 | 88.26±5.71 | 0.9308±0.0322 |
| ResNet-50 | Supervised | Random | 78.61±4.91 | 86.47±3.58 | 87.07±3.17 | 86.34±5.36 | 87.24±3.81 | 0.9231±0.0220 |
| ResNet-101 | Supervised | Random | 82.39±5.11 | 90.58±3.53 | 91.35±2.84 | 90.43±5.55 | 91.38±3.64 | 0.9482±0.0196 |
| EfficientNet-B7 | Supervised | ImageNet | 83.09±5.93 | 91.11±2.91 | 91.06±2.78 | 89.52±4.82 | 93.42±3.11 | 0.9698±0.0113 |
| VGG19 | SSL | CR | 82.13±5.73 | 89.02±4.18 | 89.69±3.22 | 87.52±5.97 | 91.24±4.37 | 0.9543±0.0237 |
| ResNet-50 | SSL | CR | 81.17±5.66 | 88.46±5.57 | 89.76±2.76 | 86.14±6.37 | 91.59±6.51 | 0.9539±0.0260 |
| ResNet-101 | SSL | CR | 83.83±5.79 | 90.49±3.57 | 90.89±3.57 | 89.92±5.57 | 91.71±3.69 | 0.9557±0.0256 |
| VGG19 | SSL | SimCLR | 82.50±3.39 | 89.24±2.04 | 90.81±2.66 | 88.26±3.95 | 90.89±2.30 | 0.9644±0.0248 |
| ResNet-50 | SSL | SimCLR | 81.86±4.43 | 91.06±2.80 | 91.71±2.71 | 89.78±5.83 | 93.03±2.68 | 0.9623±0.0161 |
| ResNet-101 | SSL | SimCLR | 82.28±3.39 | 91.61±3.57 | 91.39±2.37 | 90.65±4.67 | 93.25±4.24 | 0.9758±0.0183 |
| VGG19 | SSL | SimSiam | 82.55±5.62 | 90.59±3.39 | 91.07±3.36 | 90.28±4.32 | 91.55±4.21 | 0.9716±0.0202 |
| ResNet-50 | SSL | SimSiam | 82.03±4.29 | 90.32±4.11 | 90.29±2.59 | 89.98±4.65 | 91.31±5.36 | 0.9616±0.0137 |
| ResNet-101 | SSL | SimSiam | 83.97±4.98 | 91.80±2.99 | 92.10±2.35 | 90.96±1.56 | 93.32±2.30 | 0.9771±0.0156 |
| VGG19 | SSL | CTL | 84.77±3.96 | 91.16±2.41 | 91.93±4.71 | 90.06±4.30 | 92.94±2.65 | 0.9725±0.0145 |
| ResNet-50 | SSL | CTL | 83.63±4.55 | 90.24±3.23 | 90.54±1.94 | 89.60±3.65 | 91.54±4.68 | 0.9679±0.0168 |
| ResNet-101 | SSL | CTL | **85.38±5.51**\*\* | **92.22±3.98**\* | **92.66±1.45** | **91.17±4.99** | **93.96±4.72** | **0.9798±0.0157**\* |

Init: initialization for pre-training. Random represents that the network was trained from scratch based on a Gaussian distribution, and ImageNet represents that the network was pre-trained on the ImageNet dataset. The number shown in bold indicates the best result in each column, and the underlined numbers indicate the best results of different types of baselines. The significance level for a Wilcoxon signed-rank test[37] is 0.05; *: *p*-value < 0.05; **: *p*-value < 0.01.

**Table IV** indicates that in the binary classification task, CTL on ResNet-101 also achieved the highest accuracy, sensitivity, specificity, and AUC values. It obtained a 92.66±1.45% accuracy, 0.56% higher than SimSiam on ResNet-101. Compared with the ResNet-101 network trained by supervised learning, the accuracy of our method was increased by 1.31%. Moreover, CTL on ResNet-101 got the highest AUC value of 0.9798±0.0157, 1% higher than that of EfficientNet-B7 pre-trained on the ImageNet dataset. Although transfer learning from natural image collections has been widely used in medical image analysis, the gap between natural and medical images may lower the performance of EfficientNet-B7. Generally speaking, SimSiam on ResNet-101 performed the best among the three contrastive learning strategies in the binary classification task. Compared with this best baseline, the sensitivity, specificity, and AUC values of CTL on ResNet-101 were increased by 0.27%, 0.64%, and 0.27%, respectively. Moreover, there is a significant difference between CTL and SimSiam regarding AUC. Therefore, this result further suggests that contrastive texture feature learning can help improve overall classification performance to detect high-risk cervical lesions more accurately.

3.2 Machine-machine comparison for OCT patches

TABLE V. COMPARISON AMONG MEDICAL EXPERTS AND OUR METHOD (95% CI)

| Model&Experts | Five-class classification | | Binary classification | | | | |
|---|---|---|---|---|---|---|---|
| | *Accuracy* (%) | *F1-score* (%) | *Accuracy* (%) | *Sensitivity* (%) | *Specificity* (%) | PPV (%) | NPV (%) |
| Investigator1 | **91.58** | **95.61** | **98.87** | 97.55 | **99.89** | **99.86** | 98.14 |
| | (90.58–92.51) | | (98.45–99.20) | (96.61–98.29) | (99.61–99.99) | (99.48–99.98) | (97.42–98.70) |
| Investigator2 | 86.24 | 92.61 | 91.98 | 81.95 | 99.73 | 99.57 | 87.73 |
| | (85.01–87.40) | | (90.99–92.88) | (79.85–83.91) | (99.37–99.91) | (99.10–99.86) | (86.25–89.10) |
| Investigator3 | 88.22 | 93.74 | 93.90 | 91.74 | 95.57 | 94.11 | 93.74 |
| | (87.07–89.31) | | (93.02–94.69) | (90.19–93.2) | (94.52–96.46) | (92.75–95.29) | (92.55–94.79) |
| Investigator4 | 87.95 | 93.59 | 97.13 | 96.16 | 97.89 | 97.24 | 97.06 |
| | (86.79–89.05) | | (96.50–97.68) | (96.11–97.09) | (97.13–98.50) | (96.25–98.03) | (96.18–97.77) |
| Avg. (95% CI) | 88.50* | 93.90 | 95.45 | 91.85 | 98.27*** | 97.70 | 94.17 |
| | (87.36–89.57) | | (94.70–96.16) | (90.30–93.24) | (97.57–98.81) | (96.66–98.37) | (92.82–94.99) |
| Ours (95% CI) | 87.16 | 93.14○ | 95.21 | **98.53*** | 92.64 | 91.19 | **98.79** |
| | (85.96–88.28) | | (94.42–95.92) | (97.76–99.09) | (91.36–93.79) | (89.67–92.56) | (98.15–99.25) |

PPV: positive predictive value (*TP*/(*TP+FP*)). NPV: negative predictive value (*TN*/(*TN+FN*)). CI: confidence interval. CIs for accuracy, sensitivity, specificity, PPV, and NPV are exact Clopper-Pearson confidence intervals at the 95% confidence level.[38] The bold number indicates the best result in each column, and the underlined number indicates the best result of the investigators. The significance level for a Wilcoxon signed-rank test[37] is 0.05; ○: $p$-value > 0.05; *: $p$-value < 0.05; **: $p$-value < 0.01; ***: $p$-value < 0.001.

**Table V** presents the comparison results between CTL on ResNet-101 and four investigators on the test set in the two classification tasks. In the five-class classification task, our method's accuracy (F1-score) was 1.34% (0.76%) lower than the average of the four investigators. There is no statistically significant difference between them regarding F1-score. Our method achieved almost the same accuracy value (95.21% vs. 95.45%) as the average of the investigators in the binary classification task. Compared with Investigator1, the best performer, our method's sensitivity and NPV values were increased by 0.98% and 0.65%, respectively. A significant difference occurs between our method and the average of investigators regarding sensitivity. This result indicates that our method can correctly identify high-risk cervical diseases in OCT images. However, our method's specificity and PPV values were 5.63% and 6.51% lower than the average of the investigators, respectively, suggesting much room for improvement.

**Figure 3** displays the confusion matrices of our method and the four investigators in the human-machine comparison experiment. In the five-class classification task, most human misclassifications occurred between HSIL and cervical cancer. Besides, the second and third investigators misclassified many HSIL samples into inflammation. Unlike the investigators, our method made most misclassifications in identifying cyst and inflammation correctly. In the binary classification task, our method's false-positive rate was higher than those of the investigators. The main reason is that a few OCT patches of inflammation and cyst were misclassified by our method as HSIL and cancer, possibly due to similar texture features that are hard to discern. Instead, the investigators' false-negative rates were higher than that of our method. For example, the best investigator's misclassification rate of high-risk was increased by 9.80% compared with our method. The main reason is that the investigators reviewed patches rather than the entire OCT volume so that they tended to make relatively conservative diagnoses, especially for cervical cancer. Generally speaking, our method reached the average level of the four investigators and can provide better diagnostic results than the second and third investigators in the binary classification task (see **Figure 4**).

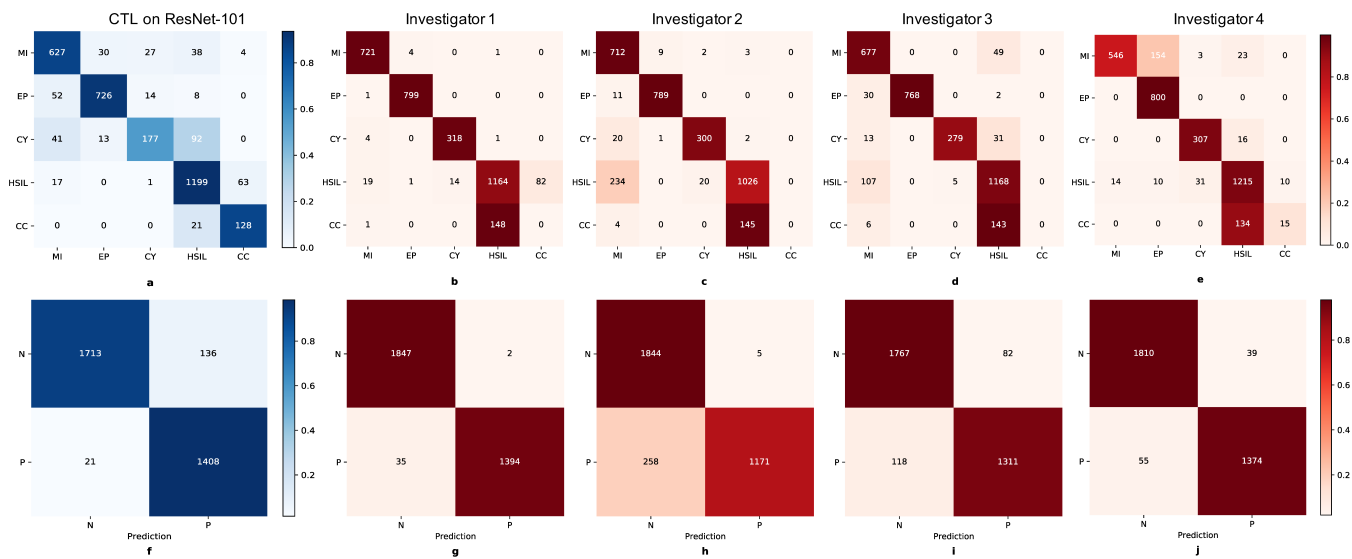

Fig. 3.  Confusion matrices of our method and the four investigators. (a)–(e): the five-class classification task; (f)–(j): the binary classification task.

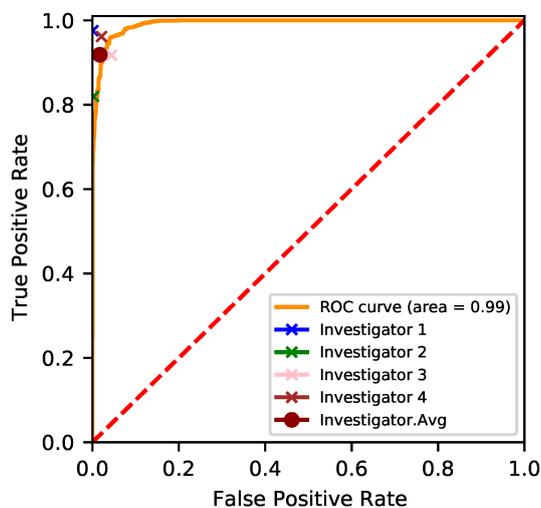

Fig. 4.  A ROC curve for the binary classification task.

3.3 External validation for OCT volumes

To evaluate the generalization of our CADx method, we experimented on a different dataset, including 287 cervical OCT volumes from another hospital, a fifth of which were of high risk. Our method outputs a binary classification result for the input OCT volume using the cross-shaped threshold voting strategy based on patches. **Table VI** shows the classification results on the external validation and test sets. The two datasets are of similar size.

TABLE VI. VOLUME-LEVEL BINARY CLASSIFICATION RESULTS OF OUR METHOD AND MEDICAL EXPERTS (95% CI)

| Subject | Dataset | *Accuracy* (%) | *Sensitivity* (%) | *Specificity* (%) | PPV (%) | NPV (%) | *F*1-score (%) |
|---|---|---|---|---|---|---|---|
| CTL on ResNet-101 | Test | 95.39 | 96.36 | 95.06 | 86.89 | 98.72 | 95.71 |
| | | (91.69–97.77) | (87.47–99.56) | (90.50–97.84) | (75.78–94.16) | (95.45–99.84) | |
| CTL on ResNet-101 | External validation | 96.17 | **91.53** | 97.37 | 90.00 | **97.80** | 94.36** |
| | | (93.25–98.07) | (81.32–97.19) | (94.36–99.03) | (79.49–96.24) | (94.93–99.28) | |
| Investigator1 | External validation | **97.91** | 90.32 | **100** | **100** | 97.40 | **94.92** |
| | | (95.51–99.23) | (80.12–96.37) | (98.37–100) | (93.62–100) | (94.43–99.04) | |
| Investigator2 | External validation | 95.47 | 90.32 | 96.89 | 88.89 | 97.32 | 93.49 |
| | | (92.38–97.57) | (80.12–96.37) | (93.70–98.74) | (78.44–95.41) | (94.26–99.01) | |
| Investigator3 | External validation | 95.47 | 85.48 | 98.22 | 92.98 | 96.09 | 91.41 |
| | | (92.38–97.57) | (74.22–93.14) | (95.51–99.51) | (83.00–98.05) | (92.70–98.20) | |
| Investigator4 | External validation | 94.77 | 80.65 | 98.67 | 94.34 | 94.87 | 88.75 |
| | | (91.53–97.05) | (68.63–89.58) | (96.15–99.72) | (84.34–98.82) | (91.21–97.32) | |
| Avg. (95% CI) | External validation | 95.91 | 86.69 | 98.44 | 93.89 | 96.41 | 92.20 |
| | | (92.92–97.88) | (75.66–93.98) | (95.83–99.62) | (84.24–98.50) | (93.11–98.41) | |

As shown in **Table VI**, our CADx method achieved similar volume-level classification results (in terms of accuracy, sensitivity, specificity, and F1-score) on the two datasets from different hospitals, indicating a good capability of robustness and generalization for female patients from different regions. Besides, it is worth noting that the sensitivity value on the external validation set was decreased by 4.83%, suggesting further improvement on our method's ability to deal with intractable cervical lesions in OCT images.

In the volume-level binary classification task, our method achieved an advantage of 0.26% (2.16%) higher than the four investigators' average accuracy (F1-score). It obtained an accuracy (F1-score) value comparable to that of the first investigator, the best performer in this task. Compared with Investigator1, our method's sensitivity and NPV values were increased by 1.21% and 0.40%, respectively. Although **Table V** shows that there is still a gap between our method and Investigator1 in specificity and PPV, our method's specificity and PPV values are close to the average levels of the investigators. Moreover, a significant difference occurs between our method and the average of investigators in the overall classification performance (F1-score). Hence, the results indicate that our method can work like skilled medical experts, with an average-to-good diagnostic capability.

3.4 Feature visualization

In addition to the heat matrix for cervical OCT volumes that assist in localizing high-risk lesions, we also used CAM to visualize the learned image features to help gynecologists better interpret the patch-level diagnoses made by our CADx method. **Figure 5** displays six typical cases that belong to ectropion, HSIL, and cervical cancer, and each case contains three types of images: OCT, LBP, and CAM.

The cases of cervical ectropion are placed in the first row of **Figure 5**. Although cervical ectropion is a non-cancerous condition for women, sometimes the OCT images of EP are often misclassified as HSIL or cervical cancer by inexperienced gynecologists or pathologists. The OCT images in **Figures 5a** and **5b** revealed no layered structure and the papillary structure (PS) with hyper-scattering boundaries in ectropion tissues (see the red arrows). The CAM heat maps in the first row of **Figure 5** highlighted the papillary structures and interpapillary ridges in the corresponding OCT images, which is consistent with the histomorphological finding in previous studies.[5, 6, 8] Besides, the LBP maps captured sharp contours of ectropion tissues as a noticeable texture feature.

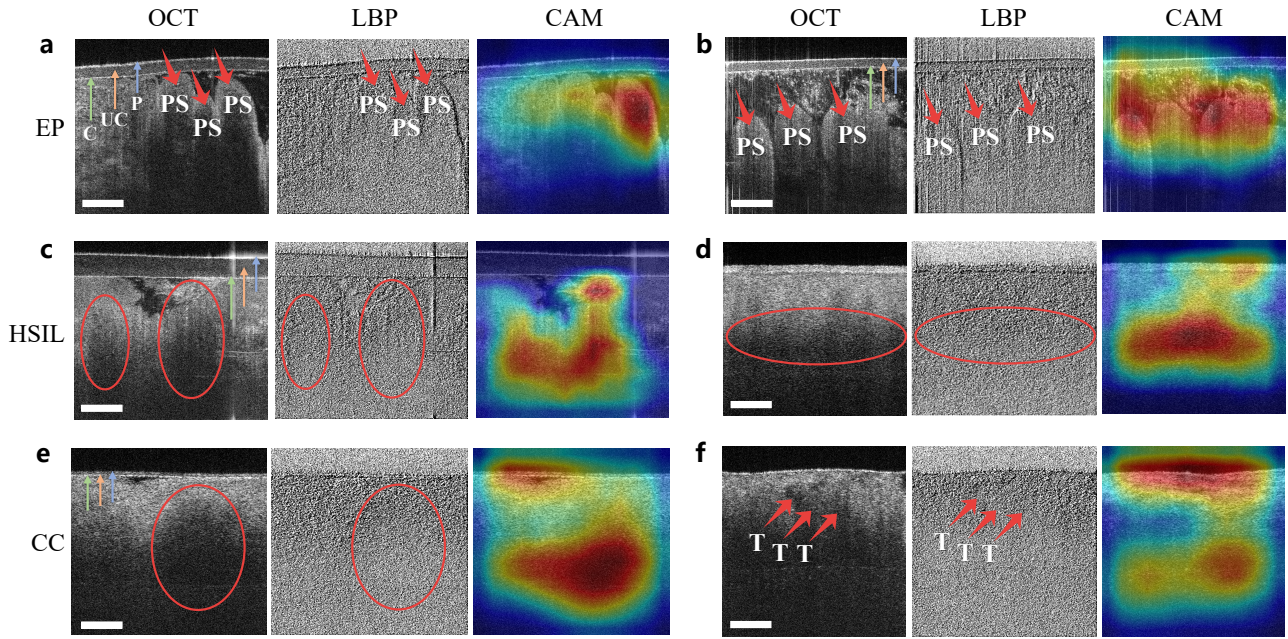

Fig. 5. Pixel-level histomorphological feature visualization. (a) and (b): cervical ectropion; (c) and (d): HSIL; (e) and (f): cervical cancer: squamous cell carcinoma (SCC). PS: papillary structures with hyper-scattering boundaries; T: tumor; C: condom; UC: ultrasound couplants; P: probe. Scale bars: 200 μm.

The OCT images in the second row of **Figure 5** demonstrated two cases of HSIL, where we cannot find the typical layered structure and the basement membrane. Moreover, the image intensity decayed rapidly as the tissue depth increased, thus forming hyper-scattering icicle-like shapes (see the regions surrounded by the two red eclipses in **Figure 5c**).[5] The OCT image in **Figure 5d** further presented distinct patterns of light and dark composed of such icicle-like shapes (see the region surrounded by the red eclipse), and the stroma below was no longer observed. The CAM heat maps highlighted these histomorphological features[5] in the corresponding OCT images. Unfortunately, the LBP maps show little human-readable texture information to help interpret HSIL, which deserves further investigation.

Two cases of cervical cancer (more specifically, squamous cell carcinoma (SCC)) are placed in the third row of **Figure 5**. Because the standard squamous epithelial structure and basal membrane were lost entirely, the OCT images revealed a complete lack of architectural polarity.[6, 8] Although the OCT image in **Figure 5e** shows a pattern similar to HSIL, the image intensity decayed more rapidly. It formed hypo-scattering regions surrounded by the red eclipse. In a physical sense, the high density of tumor cells affects light penetration in tumor tissue. The OCT image in **Figure 5f** presents another pattern, i.e., several irregular nests of heterogeneous regions composed of epithelial and tumor cells (see the red arrows) were detected in the epithelial layer.

Although the LBP maps provided little explainable evidence for diagnosing SCC, the CAM heat maps highlighted the histomorphological features of SCC in OCT images recognized in previous studies.[5, 8]

## 4 DISCUSSION

Since OCT is an emerging biomedical imaging technology, most of the current work in gynecology is to assess the clinical effectiveness of OCT in manually diagnosing high-risk cervical lesions.[5, 8-10] Due to the lack of licensed OCT devices to acquire OCT images from the cervix, there are no publicly available cervical OCT datasets for researchers to train image classification models. CNN-based classification models have been proven to perform better than traditional machine learning algorithms in natural and medical image analysis, but they require massive amounts of labeled images. Compared with numerous deep-learning-based classification models for cervical cytology, colposcopy, and histopathology images, few CNN-based cervical OCT image classification models are available for comparison.[39-42] This study aims to propose a new SSL strategy for efficient OCT image classification model training rather than to design a new classification model. To this end, we selected CNN benchmarks as backbone networks for the downstream classification task for a fair comparison.

### 4.1 Sensitivity analysis of LBP parameters

We ablated LBP parameters (i.e., $R$ and $P$) to analyze the parameter sensitivity of LBP. We sampled a subset from the cervical OCT image set, where each class had the same number of OCT images to avoid the class imbalance problem. The backbone network was ResNet-101. For consistency, we used the same settings as the machine-machine comparison experiment. **Figure 6** presents the classification results of ResNet-101 pre-trained by CTL with different LBP parameter values for 30 epochs. It is evident from **Figure 6** that the accuracy value reaches the maximum when $R$ and $P$ are equal to 4 and 32, respectively. Therefore, the values of $R$ and $P$ were set to 4 and 32 in this study's experiments.

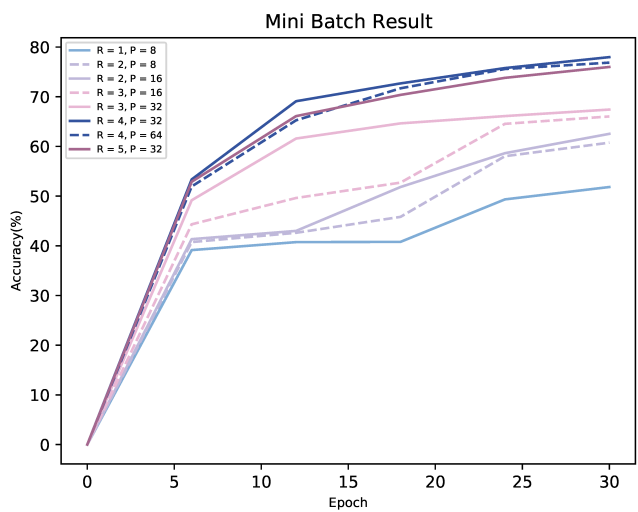

Fig. 6. Sensitivity analysis of LBP parameters on classification performance.

4.2 Size effect of labeled data

Despite the utilization of SSL, we also experimented with four different sizes of labeled data, i.e., 25%, 50%, 75%, and 100% of the patch data used for the downstream classification task, to evaluate the impact of labeled data size on our method's prediction performance. Due to limited image data and computing resources, we selected ResNet-101 instead of complex DenseNet[43] as the backbone network for this experiment. Also, we did not use a large batch size like 256 for model training, and the batch size was still set to 32. **Table VII** presents the ten-fold cross-validation results of the backbone network pre-trained with different data sizes.

TABLE VII. TRAINING ON DIFFERENT SIZES OF LABELED DATA

| Size | Classifier | *Accuracy* (%) | *F*1-*score* (%) | AUC |
|---|---|---|---|---|
| 25% | ResNet-101 (Random) | 53.23±5.58 | 64.21±4.67 | 0.7974±0.0216 |
|  | CR on ResNet-101 | 58.91±5.29 | 67.49±4.32 | 0.8259±0.0184 |
|  | SimCLR on ResNet-101 | 62.73±4.79 | 74.73±3.96 | 0.8733±0.0468 |
|  | SimSiam on ResNet-101 | 64.15±4.39 | 78.62±3.73 | 0.8824±0.0358 |
|  | CTL on ResNet-101 | **66.69±3.42** | **80.32±3.13** | **0.8911±0.0289** |
| 50% | ResNet-101 (Random) | 76.36±4.74 | 83.29±4.21 | 0.9125±0.0144 |
|  | CR on ResNet-101 | 78.29±4.17 | 84.13±3.94 | 0.9235±0.0182 |
|  | SimCLR on ResNet-101 | 81.14±4.07 | 84.92±3.17 | 0.9381±0.0173 |
|  | SimSiam on ResNet-101 | 81.39±3.91 | 86.63±4.82 | 0.9437±0.0168 |
|  | CTL on ResNet-101 | **82.25±3.93** | **90.82±4.93** | **0.9575±0.0124** |
| 75% | ResNet-101 (Random) | 78.54±4.53 | 87.36±5.06 | 0.9381±0.0106 |
|  | CR on ResNet-101 | 81.62±3.97 | 89.21±3.27 | 0.9485±0.0113 |
|  | SimCLR on ResNet-101 | 82.13±3.65 | 89.78±3.18 | 0.9517±0.0148 |
|  | SimSiam on ResNet-101 | 82.96±3.59 | 91.69±4.32 | 0.9588±0.0203 |
|  | CTL on ResNet-101 | **83.58±2.50** | **91.74±3.36** | **0.9693±0.0156** |
| 100% | ResNet-101 (Random) | 82.39±5.11 | 90.58±3.53 | 0.9482±0.0196 |
|  | CR on ResNet-101 | 83.83±5.79 | 90.49±3.57 | 0.9557±0.0256 |
|  | SimCLR on ResNet-101 | 82.28±3.39 | 91.61±3.57 | 0.9758±0.0183 |
|  | SimSiam on ResNet-101 | 83.97±4.98 | 91.80±2.99 | 0.9771±0.0156 |
|  | CTL on ResNet-101 | **85.38±5.51** | **92.22±3.98** | **0.9798±0.0157** |

Accuracy is used for the five-class classification task.
The bold number indicates the best result in each group of data sizes.

ResNet-101 with random initialization performed worse in case of insufficient labeled data. For example, in the five-class classification task, ResNet-101 trained by half of the labeled data obtained a 76.36±4.74% accuracy (83.29±4.21% F1-score). Compared with the one trained by all the labeled data, the accuracy (F1-score) value was decreased by ~6% (7.29%), and the AUC value was decreased by 3.57% in the binary classification task. Instead, with the help of contrastive learning (i.e., CTL, CR, SimCLR, and SimSiam), the network's performance can be visibly improved on small-scale training datasets. In this case, CR performed worse than the other three strategies, perhaps due to random noises caused by the image restoration based on patch selection and swap. For example, when using 50% of labeled data, the accuracy (F1-score) values of ResNet-101 pre-trained with CTL and SimSiam were improved by 5.89% (7.53%) and 5.03% (3.34%), respectively, compared with that of ResNet-101 with random initialization. Moreover, we can see more significant improvement from **Table VII** when only 25% (or less) of labeled data is available. Therefore, the CTL strategy can help further improve the performance of CNNs for cervical OCT image classification, especially on large-scale datasets of OCT volumes without manual labels.

4.3 Misclassification analysis

In the five-class classification task, the misclassifications made by the four investigators occurred mainly between HSIL and CC, though these two types are high risk. Instead, CTL on ResNet-101 was able to identify them more accurately than the investigators. The reasons for this result are two-fold. First, when the investigators read a cervical OCT patch, they could not exploit the target patch's critical spatial and semantic correlations with the other ones in the same frame. Hence, the investigators tended to make relatively conservative diagnoses according to the local information available. Second, it is not easy for the investigators to distinguish CC from HSIL in visually similar features of the optical intensity distribution in OCT images (see the examples shown in **Figures 5c** and **5e**). Our method made fewer misclassifications of CC into HSIL than the investigators by considering the LBP texture.

Here, we introduce a similarity index between two patches based on the LBP texture that presents little human-readable information. Suppose $\mathbf{x}_i$ and $\mathbf{x}_j$ denote two vectors of texture features extracted from patches $i$ and $j$ by LBP. The patch similarity is defined based on the cosine similarity as

$$sim(\mathbf{x}_i, \mathbf{x}_j) = \frac{\mathbf{x}_i \cdot \mathbf{x}_j}{\|\mathbf{x}_i\| \|\mathbf{x}_j\|}, \qquad (12)$$

where $\|\mathbf{x}\|$ is the norm of vector $\mathbf{x}$.

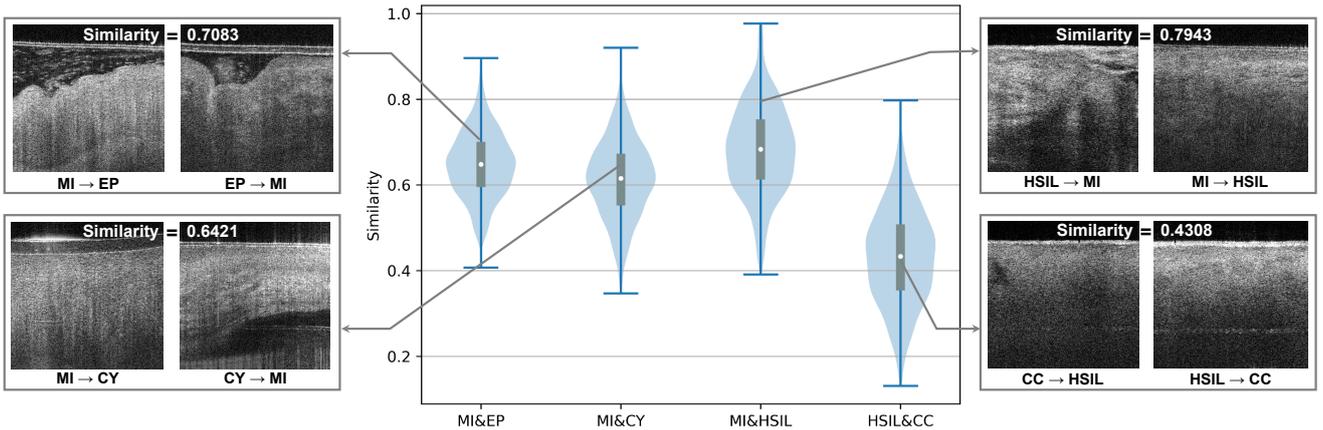

Fig. 7. Violin plots for the similarity distributions of patches misclassified by our method (the first three columns) and the investigators (the last column).

**Figure 7** uses standard violin plots to display the similarity distributions of patches misclassified by our method and the investigators. For those OCT patches with a CC (or HSIL) label misclassified into HSIL (or CC) by the investigators, the median score of their patch similarities (see the white circle in the last column) is low (< 0.5), suggesting that our method has a good capability of distinguishing between the two high-risk cervical lesions by contrastive texture learning. Compared with the investigators, our method had a higher false-positive error rate. For example, it misclassified 38 patches with an MI label into HSIL (see **Figure 3a**), possibly due to similar LBP texture features. It is evident from **Figure 7** that the median similarity score of MI and HSIL patches reaches almost 0.7. Besides, for low-risk cervical OCT patches, misclassifications made by our method occurred between MI and EP and between MI and CY. In these two cases, the median similarity scores were relatively high (> 0.6), implying that they also shared similar LBP texture features. Therefore, one of our future works is to improve the texture extraction algorithm according to the optical intensity distribution in cervical OCT images.

## 4.4 Limitation and future work

Low-grade squamous intraepithelial lesion (LSIL), previously known as CIN I, is a low-risk cervical disease. Due to a small number of recruited patients with LSIL, OCT images from LSIL tissues were excluded from the OCT image dataset used in this study. We will update this dataset by adding *in-vivo* OCT volumes of LSIL in the future. Different attention mechanisms have been recently utilized to improve the backbone network's performance for medical images.[44] Therefore, we plan to introduce specific attention mechanisms such as spatial and channel-wise attention[45] and self-attention[46] to the classification model. Finally, although we design an explainable voting strategy for cervical OCT volumes, it also has room for improvement from the following two aspects.

First, we ignore that a cervical OCT volume may have more than one type label in this study. For example, a cervical specimen often contains an inflammation of the cervix and small cysts. Therefore, we plan to introduce the multi-label classification technique[47] to the classification model of our future work. Second, since the cross-shaped threshold voting strategy only outputs a binary result, we will design a trainable voter using efficient machine learning algorithms to better aggregate patch-level prediction results with a set of labels.

## 5 CONCLUSIONS

We developed a CADx approach based on SSL for *in-vivo* 3D OCT volumes from the cervix. In particular, we proposed a contrastive learning strategy based on texture features to pre-train CNN models for cervical OCT image classification. According to the ten-fold cross-validation results on an OCT image dataset from a multi-center clinic study, our method outperformed other competitive models pre-trained from scratch, on the ImageNet dataset, and by a state-of-the-art contrastive learning method (i.e., SimSiam), in the five-class and binary classification tasks. Moreover, it performed better than two out of four medical experts on the test set in the binary classification task. Besides, we demonstrated the robustness and generalization of our method for high-risk OCT volumes on the test and external validation sets. Due to better interpretability based on texture features, the proposed CADx approach holds great potential to help gynecologists make rapid and accurate diagnoses in the clinical protocol of "see-and-treat."

ACKNOWLEDGMENT

This work was supported by the Science and Technology Major Project of Hubei Province in China (Next-Generation AI Technologies) under Grant No. 2019AEA170. In addition, the authors appreciate Tao Xu, Hao Hu, and Di Meng for their valuable work of OCT image collection.

CONFLICT OF INTEREST STATEMENT

The authors have no relevant conflicts of interest to disclose.